\documentstyle[prd,aps,eqsecnum,tighten]{revtex}
\begin{document}
\draft
\title{On Principle of Universality of Gravitational Interactions}
\author{ I.~B.~Pestov\thanks{Electronic address:  pestov@thsun1.jinr.ru}}
\address{Bogoliubov Laboratory of Theoretical Physics, Joint
Institute for Nuclear Research \\ Dubna, 141980, Russia}

\date{\today}
\maketitle
\begin{abstract}

In this work, the experiment is discussed on the verification of the
principle of universality of gravitational interactions and some related
problems of gravity theory and physics of elementary particles. The meaning
of this proposal lies in the fact that the self-consistency of General
Relativity, as it turns out, presuppose the existence of the nongravitating
form of energy. Theory predicts that electrons are particles that transfer
the nongravitating form of energy.

\end{abstract}

\section{Introduction}
The modern stage of development of gravity theory is characterized not only
by the search for new effects and new experiments but also by a deeper
analysis of fundamentals of the theory and conceptual problems including
an important problem of the energy of gravitational field [1,2,3].
Difficulties caused by the nontensor character of quantities describing
the energy of gravitational field turn out to be so serious that they
are considered as manifestation of specific properties of the gravitational
field: universality, nonscreening nature, nonlocalizability. A detailed
analysis shows that none of specific properties of gravitational field
can explain the so-called nonlocalizability of that field. Not only the
energy but also all the results of theory, except for the Lagrange function
and gravitational field equations, appear to be noncovariant. So, in
general relativity, a nonstandard situation occurred: in the theory
whose principles are mathematically formulated rigorously, important
physical consequences are in contradiction with the initial statements.

So, when general relativity is formulated, a general logical requirement
admissibility of arbitrary systems of coordinates is postulated, however,
it turns out that in the constructed theory, the dynamic characteristics of the
gravitational field (except for the Einstein equations), the density of
energy and momentum, are described by nontensor quantities. As a result,
it is impossible to uniquely describe the distribution of energy-momentum
of any physical system in the gravitational field. Therefore, there occurs
the notion of nonlocalizability of the gravitational field. The energy of
this field is not localizable, i.e. there is no uniquely defined energy
density.

\section{Nonlocalizability}
The nature of this phenomenon is as follows. While the electromagnetic
field is described both by a vector potential and a metric, the Einstein
law of gravity [4] does not contain anything except gravitational
potentials.  For the electromagnetic field, the physical quantity is
the class of equivalence of vector potentials determined by one
arbitrary function.  A representative of each class of equivalence is
chosen by imposing the Lorentz condition that is generally covariant,
i.e. independent of a particular system of coordinates, since the
theory contains the so-called background object, the Minkowski
metric. At the same time, the physical quantity for gravitational
field is the class of equivalence of gravitational potentials defined
by four arbitrary functions. Only one quantity, the Einstein-Hilbert
action, is independent of the choice of these functions. Extending
analogy, we note that different representatives of the class of
equivalence in the Maxwell theory correspond not only to the same
action but also to the so-called tensor of electromagnetic field.
Therefore, the same Lorentz force and energy density correspond to
different representatives of the equivalence class.  In the Einstein
theory, different representatives of the equivalence class correspond
to the same gravitational field that is differently oriented in
space-time with respect to the same observer.  Different
representatives correspond to different orientations. Ambiguity in
the choice of orientation is determined by four arbitrary functions
of coordinates. Since the theory does not contain any objects besides
gravitational potentials, the representative from each equivalence class
can be chosen in a general covariant manner only through introducing, into
the theory, the nondynamic so-called background object, background
metric [5].

The choice of a representative from each equivalence class is achieved by
imposing four general covariant conditions on covariant derivatives of
gravitational potentials with respect to the background connection.
Nonlocalizability of the gravitational field is then defined by the
freedom in choice of the background metric or background connection. Thus,
the problem of energy of the gravitational field is reduced to the problem
of physical meaning of the background connection that becomes of principal
importance. If the gravitational field and background connection are
given on the same manifold, gravitating particles are moving along
geodesics defined by gravitational potentials. Then, a natural question
arises concerning the nature of particles moving along geodesics given
by the background connection. Existence of particles of that sort is
an evident necessity without which it is a difficult problem to discuss the
physical meaning of the background connection. The latter problem can be
avoided by assuming that the particles are moving along geodesics of the
background connection when there is no gravitational field. Then it
follows that the background connection has the meaning only in the absence
of gravitational field.

So, based on purely logical requirements following
>from known facts, we conclude that there exists nongravitating form of the
energy that is directly related to the physical meaning of gravitational
potentials in the framework of general covariance principle. As it has
already been underlined, the latter is a purely logical requirement
inherent in any physical theory, including gravity theory. Hence it
becomes necessary to experimental verify the universality principle of
gravitational interactions. "The validity of this principle in the
microscopic domain is not as evident as the validity of the
coincidence axiom. We know of many rules which apply with great
rigor to electromagnetic and other types of interaction and it is
conceivable that the special role of the gravitational
interaction may dissolve in higher harmony. For this reason, I
shall pay prime attention to Einstein`s first observation, that only
coincidences have a direct physical meaning, values of coordinates do
not."  This citation is taken from paper by Wigner [6] in order to
emphasize that it is important to test the universality principle of
gravitational interactions not only for the problem of
self-consistency of general relativity but also for explanation of
the role of gravitational forces in the physics of microworld.

\section{On gravity law}
We will mathematically formulate main distinctive features of general
relativity on the basis of the Einstein gravity law [4]. This law,
as viewed in the framework of mathematical analysis, is the following system
of nonlinear second-order partial differential equations
\begin{equation}
			R_{ij} = 0
\end{equation}
for ten functions $g_{ij}(x)$ of 4 independent variables $x^0, x^1, x^2,
x^3$. This system of equations possesses the following remarkable
property.  Let the functions $g_{ij}(x)$  have a common domain of
definition $D$ and are a solution to system (1). Now, consider
functions  $\varphi^i(x), \quad i=0,1,2,3,$ such that their domain of
definition and range of their values contain $D$, and the Jacobian
$J = |\partial\varphi^i/\partial{x^j}|$ differs from zero in $D$. As
it is known, in this case in the domain $D$ there exist functions
$f^i(x),$  such that
$$\varphi^i(f(x)) = x^i, \quad f^i(\varphi(x)) = x^i.$$
Next, let us constitute the functions
\begin{equation}
	       \tilde g_{ij}(x) = g_{kl}(f(x)) f^k_i (x) f^l_j(x),
\end{equation}

where  $f^k_i(x) = \partial_if^k(x).$
Substitution of (2) into (1) shows that the
functions $ \tilde g_{ij}(x)$   are a new solution to equations (1)
provided that the functions $g_{ij}(x)$ in (2) are a solution to those
equations. This analytic aspect of the gravity law is a crucial
point. We will now consider its manifestation in various problems.

First, we shall examine the very important Cauchy problem for eqs. (1).
The metrics $g_{ij}(x)$ and $ \tilde g_{ij}(x)$ , as mentioned
above, describe the same physical situation. It is associated with
the whole class of equivalence of solutions to eqs. (1) determined by
4 functions. To choose a certain element from every class of
equivalence, it is convenient to introduce a global "background"
metric and to impose 4 conditions on covariant derivatives of the
physical metric with respect to the Levi-Civita background metric
$\hat g_{ij}$ [8]  $$\hat \nabla_i (\sqrt{-g} g^{ij}) = 0,$$
which remove arbitrariness defined by eqs.(2). The result is reduced
equations of the hyperbolic type for the metric $g_{ij}$   with respect to
the global background metric $\hat g_{ij}.$ The background metric is quite
necessary for writing "gauge conditions" in a general-covariant form
independent of the choice of a coordinate system. So, to derive a
definite solution to the Einstein equations (1), one should eliminate
arbitrariness given by analytic relations (2), which is achieved as
indicated above.  Consequently, to obtain physical solutions to
equations of the gravitational field, they should be treated as a
nonlinear system of second-order equations on the manifold $M$ with
metric $\hat g_{ij}$ given on the manifold globally [6].  Note that
the notion of a continuous 4-dimensional manifold provided an
effective remedy used in considerations in modern physics.  The very
structure of manifold with its topology remains arbitrary and should
be determined by considerations, generally speaking, outside of the
scope of general relativity. Thus, the Cauchy problem in general
relativity can be brought into the general-covariant form necessary
for any physical theory only when the background connection is
introduced.

For further clearer illustration, let us trace analogy with the theory
of gauge fields. According to Einstein, the gravitational field is put
into correspondence with a symmetric tensor field of the second rank
$g_{ij},$ satisfying the nonlinear equations (1). The electrotonic
state of electromagnetic field introduced by Faraday is described by
the vector potential (1-form) $A = A_idx^i. $   The gauge transformations
$$ A \rightarrow {\bar A } = A + d \varphi $$
correspond to transformations (2). Principal difference of the gravitational
field from the electromagnetic field is that one cannot construct quantities
invariant under transformations (2) from the Einstein gravitational
potentials, whereas from components of the vector potential, the
gauge-invariant tensor of electromagnetic field (2-form) $F = dA $
can be constructed.  In particular, with the help of (2) it is not
difficult to verify that the tensor of Riemannian curvature that is
in a sense analogous to the tensor of electromagnetic field is not
invariant under the transformations of gravitational potentials (2).
The case of non-Abelian gauge fields, in the sense of existence of
gauge-invariant quantities, is closer to general relativity as
compared to the Maxwell theory. The reason is that the strength
tensor of a non-Abelian gauge field is not a gauge-invariant quantity
[10]; only the energy-momentum tensor and Lagrange function remain
gauge-invariant quantities. This can be observed as follows:  with a
non-Abelian gauge group, one cannot connect conserved gauge-invariant
"charges" analogous to the electron charge. This is a principal
difference of the Yang-Mills field theory from electrodynamics, and
just in this it is closer to general relativity, where also one
cannot introduce an invariant gravitational "charge". It is well-known
which difficulties are connected with the extension of gauge symmetry
and how they were overcome. A rather long way was required to construct
physically acceptable models with non-Abelian gauge fields. Spontaneous
breaking of symmetry as the mechanism that allows one to provide quanta
of those fields with mass and localization of interactions of the
indicated class are merely the most typical manifestations caused by
the change of the gauge group of electrodynamics to a wider gauge group
of transformations. In gravity theory, only the Einstein-Hilbert
action and equations of the gravitational field deduced from it are
invariant under transformations (2). Related difficulties are not yet
overcome. Like in the theory of non-Abelian gauge fields, in gravity
theory, one cannot introduce a gravitational charge invariant under
transformations (2).  Which are consequences and how they show
themselves physically, is still an open problem. The first important
step along this way could be the problem of existence of a
nongravitating form of energy, because having solved this problem,
one could raise the problem of physical meaning of the background
connection and analyze the corresponding consequences.  In this
connection, consider evidences for existence of a nongravitating form
of energy beyond the scope of general relativity.

\section{Nongravitating form of energy}
As it is shown in [7], the existence of nongravitating form
of energy is closely related to gauge symmetry inherent in general relativity.
The idea of symmetry like that is very simple and consists in that
coordinates on the total linear group can be put in correspondence with
tensor fields of type (1,1) on a manifold rather than a set of scalar
fields , it is accepted in gauge abstract theory. Hence, it follows that
coordinates of all subgroups of the total linear group can also be put in
correspondence with tensor fields on a manifold.This actually covers all
physically interesting groups of gauge symmetry. The gauge symmetry under
consideration is a realization of the abstract theory of gauge fields in the
framework of modern differential geometry. Its typical feature is that it
does not supposes distinction between space-time and gauge, or so-called
isotopic, space. At the same time contemporary gauge models presuppose
an exact local distinction between space-time and gauge space. Just in
this aspect, the gauge symmetry under consideration opens essentially new
possibilities. So, the total linear group and its subgroups admit a
simple realization in terms of geometrically well defined tensor fields.

The total linear group, like the group of gauge symmetry, is
invariant under transformations (2). This signifies that
transformations (2) do not break equivalence given by that gauge
group. However, any reduction of the considered gauge group to its
subgroups results in that all these subgroups are not invariant under
transformations of the symmetry group of gravitational interactions
defined locally and usually called the group of diffeomorphisms.
Thus, if at least one subgroup of the gauge group is physically
realized, the form of energy connected with this system will be
nongravitating. In the papers mentioned above it has been shown that
the Dirac theory cannot be deduced without reduction of the given
gauge group. This result is itself rather easily apprehended  within
the Dirac theory on the basis of the well-known fact that spinors are
not a basis of a representation of the total linear group considered
here as a group of transformations of coordinates. For that reason,
difficulties have arisen of fundamental character when introducing
spinors into general relativity. Strange though it may seem, the way
out of that situation was looked for not through analyzing and
verification of the universality principle of gravitational
interactions as applied to electrons but through introducing an
orthogonal basis into gravity theory. The difficulties with
geometrization of the Dirac theory were resolver in its favor.

Accurate formulation of the Dirac theory, even in the Minkowski space-time,
requires introduction of the so-called tetrad. As a result, instead of the
analysis of the Dirac theory, there appear various tetrad theories of
gravity and, which is surprising, the introduction of basis was apprehended
as a peculiar revelation not only in physics, but in geometry, as well.
However, a pure logic and evident requirement applicable to any physical
theory is the principle according to which physical laws should be
formulated in the form independent of both the choice of a coordinate
system and a basis in vector spaces associated with a given theory. The
notion of basis should be absent in the formulation of physical laws.
For the first time, this principle was formulated by Dirac who gave
a consistent construction of quantum mechanics on its basis [8].
However, it turns out that the Dirac electron theory does not the
principle he formulated. It seems that existence of the Dirac theory
very clearly points out that this theory is a reduction of a more
general theory and the simplest way to perform that reduction is to
employ an orthogonal basis. This program was realized in the papers
mentioned above. The relevant conclusion consists in that electrons
carry the nongravitating form of energy. So, the nature unambiguously
points to self-consistency of general relativity and to particles
that transfer the nongravitating form of energy, electrons.

\section{On gravitational interactions of electrons}
We will not present detailed computations, rather we dwell upon main
results following from [7]. It has been shown there that
the theory of fermion fields cannot in principle be formulated without
reduction of the total linear group as a group of gauge symmetry. As a
result, it turned out to be impossible to construct the theory invariant
under both gauge transformations and transformations of the group of
diffeomorphisms. This result could be easily understood without formulae,
as well. Reduction of the gauge group is accompanied by imposing constraints
formulated in terms of the metric tensor. Consequently, the reduced gauge
group will be invariant under transformations of the group of diffeomorphisms
if the metric tensor does not change under those transformations. According to
(2), the condition for the metric being conserved under diffeomorphisms
is expressed by the system of partial differential equations for functions...
\begin{equation}
\tilde g_{ij}(x) = g_{kl}(f(x)) f^k_i (x) f^l_j(x)  = g_{ij}(x).
\end{equation}
Equations (3) may not have solutions at all. If $g_{ij}(x)$ in (3) is
the Minkowski metric, the group of space-time symmetry that keeps the
reduced gauge group to be invariant will be the Poincare group which
is in the given case a general solution to equations (3).

When the problem of gauge-invariant determination of the energy density
was studied, it was proved that there exists canonical gauge-invariant
tensor of the energy-momentum; the latter differs from the metric tensor
of energy-momentum which is not gauge-invariant since reduction of the
gauge group breaks gauge invariance of the procedure of deriving the
metric tensor of energy momentum. As the right-hand side of the Einstein
equations contains the metric tensor of energy-momentum, it follows that
gravitational interactions cannot be introduced in a gauge-invariant way.
As noted above, the Dirac theory is realized in the reduction of the
total linear group as the group of gauge symmetry. However, the reduction
of this gauge group automatically leads to the reduction of the group
of diffeomorphisms which is a symmetry group of gravitational interactions.

Now, we will formulate conditions to hold for any gravitational-interacting
field. Let a physical object be described by an appropriate field, and
equations of the field be derived from the Lagrangian. Then the field
is gravitational-interacting provided the following condition are
fulfilled. The considered field, like the gravitational field, makes a basis
of the faithful representation of the diffeomorphism group. Varying
the action of the field with respect to the metric, we obtain the
metric tensor of energy- momentum. This symmetric tensor will obey
the well-known local conservation law on solutions to the field
equations. If the theory has any gauge symmetry, the energy-momentum
tensor should be gauge-invariant. A relevant necessary condition is
invariance of the gauge group with respect to transformations of the
group of diffeomorphisms, the symmetry group of gravitational
interactions. If the latter condition is not satisfied, there are two
possibilities. The first consists in that gauge symmetry is
eliminated from consideration, and the theory is constructed which is
invariant only with respect to transformations of the group of
diffeomorphisms. The second possibility is to construct the gauge-invariant
theory which cannot include gravitational interactions. This means that
there exists the nongravitating form of energy. As it has turned out, the
Dirac theory does not obey the formulated conditions.

The Dirac theory unambiguously points to the existence of nongravitating
form of energy and to a concrete object , the carrier of that form, the
electron. Reduction of the gauge group resulting in the Dirac wave function,
(for details see papers cited above)  reduces the group of space-time
symmetry step by step, so that the latter is to be properly restored to
ensure the Poincare invariance. This is achieved with the help of
global gauge invariance, and therefore, the electron represents the
nongravitating form of energy.

The simplest test for establishing whether the electron is that form is to
measure its weight. Experiments of that kind were performed in 1967 at
Stanford University by the Fairbank group [9]. They demonstrated a
zero result. Here we will cite the abstract of paper [9].

"A free-fall techniques has been used to measure the net vertical
component of force on electrons in vacuum enclosed by a copper tube.
This force was shown to be less than $0.09 mg$, where $m$ is the inertial
mass of the electron and $g$ is $980 cm/sec^2$. This supports the contention
that gravity induces an electric field outside a metal surface, of
magnitude and direction such that the gravitational force on electrons
is cancelled".

Thus, the explanation of experimenters of that "zero" is that gravity
induces, in an experimental setup, an electric field which just compensates
the force of gravitational attraction. It follows then that positrons under
the same conditions should fall with acceleration $2g$. Fairbank planned to
carry out experiments with positrons, however, this plan met with
difficulties. Later, when up-to-data-technologies open new possibilities,
preparation of this experiment was taken up again. Unfortunately, the
experiment with positrons remained at the level of preparation since after
1989 it was closed because of decease of the project's leader. Details of
the Fairbank experiment and subsequent discussions can be found in review
papers [10] and [11].

\section{Proposal of experiment}
What is suggested on the basis of our consideration?
To solve the problems of fundamental importance both for gravity theory
and elementary particle physics, it is necessary to perform the Fairbank
experiments both with electrons and positrons, i.e. to measure the
acceleration of electrons and positrons in the gravitational field of the
Earth. If the result is zero, in accordance with theoretical predictions,
it will first of all follow that general relativity is , like the theory
of electromagnetic fields, self-consistent. An urgent necessity will be
the problem of physically substantiated choice of the background metric and
proper violation of the symmetry of gravitational interactions, of the
physical meaning of that manifold, with which the background metric
is connected. Consideration of gravitational interactions on the
background of the Minkowski metric is quite natural.
However, other possibilities should not be neglected a priori,
especially, in connection with the problem of compatibility of gravitational
field equations and subsidiary (auxiliary?) conditions imposed on
gravitational potentials [5]. It may happen that gravitational
interactions, like weak interactions, are localized, which means that
gravitons transfer the mass. In the physics of elementary particles, it will
be necessary to consider the problem of gravitational interactions of
the emitting matter as a collective effect.

In conclusion, we note that expenditures on the experiment proposed could by
no means be compared with its fundamental importance for physics, we tried
here to demonstrate from various viewpoints.

 \begin{center}
    {References}
   \end{center}
\begin{enumerate}
\item  Einstein A. Sitzungsber. Preuss. Acad. Wiss. 1917.
\item  Hilbert D. Grundlagen der Physik, 2 Mitteilung, Gott. Nachr.
1917.
\item Landau L.D. and Lifshitz E.M. The Classical  Theory of Fields,
3rd ed. Pergamon Press, Oxford, 1971.
\item  Dirac P.A.M. General Theory Relativity. A Wiley Interscience
Publication, 1975.
\item Hawking S.W. and Ellis F.R. the Lardge  Scale Structure of
Space-Time. Cambridge University Press. 1973
\item Wigner E.P. Symmetries and Reflections. Indiana University
Press. 1970.
\item Pestov I.B.
Hadronic Journal Suppl. 1993.  v. 8, n.2. p. 99- 135.
\item  Dirac P.A.M. The Principles of Quantum Mechanics. Oxford, At
the Clarendon Press.
\item Witteborn F.C. and Fairbank W.M.
Experimental comparison of the gravitational force on freely falling
electrons and metallic electrons. // Phys.  Rev.Lett. 1967.  vol. 19,
p.1049-1052.
\item Nieto M.M. and Goldman T. The Arguments Against
"Antigravity" and the Gravitational Acceleration of Antimatter.
//Physics Reports.  1991.  v.205, n.5, p.222-281.
\item Darling
T.W., Rossi F., Opat G.I.,and Moorhead. The fall of chardged
particles under gravity: A study of experimental problems.// Rev. of
Mod.Phys. 1992. v.64, n.1, p.  237-257.

\end{enumerate}
\end{document}